\documentclass[journal=jacsat,manuscript=article]{achemso}

\usepackage[version=3]{mhchem} 

\author{Enrico Tapavicza}
\email{enrico.tapavicza@csulb.edu}
\author{Trevor Reutershan}
\author{Travis Thompson}

\affiliation[California State University, Long Beach]
{Department of Chemistry and Biochemistry, California State University, Long Beach, 1250 Bellflower Boulevard, Long Beach, CA, 90840}

\title{Ab initio simulation of the ultrafast circular dichroism spectrum of provitamin D ring-opening}

\abbreviations{IR,NMR,UV}

\begin{document}

\begin{tocentry}
\includegraphics[scale=0.095]{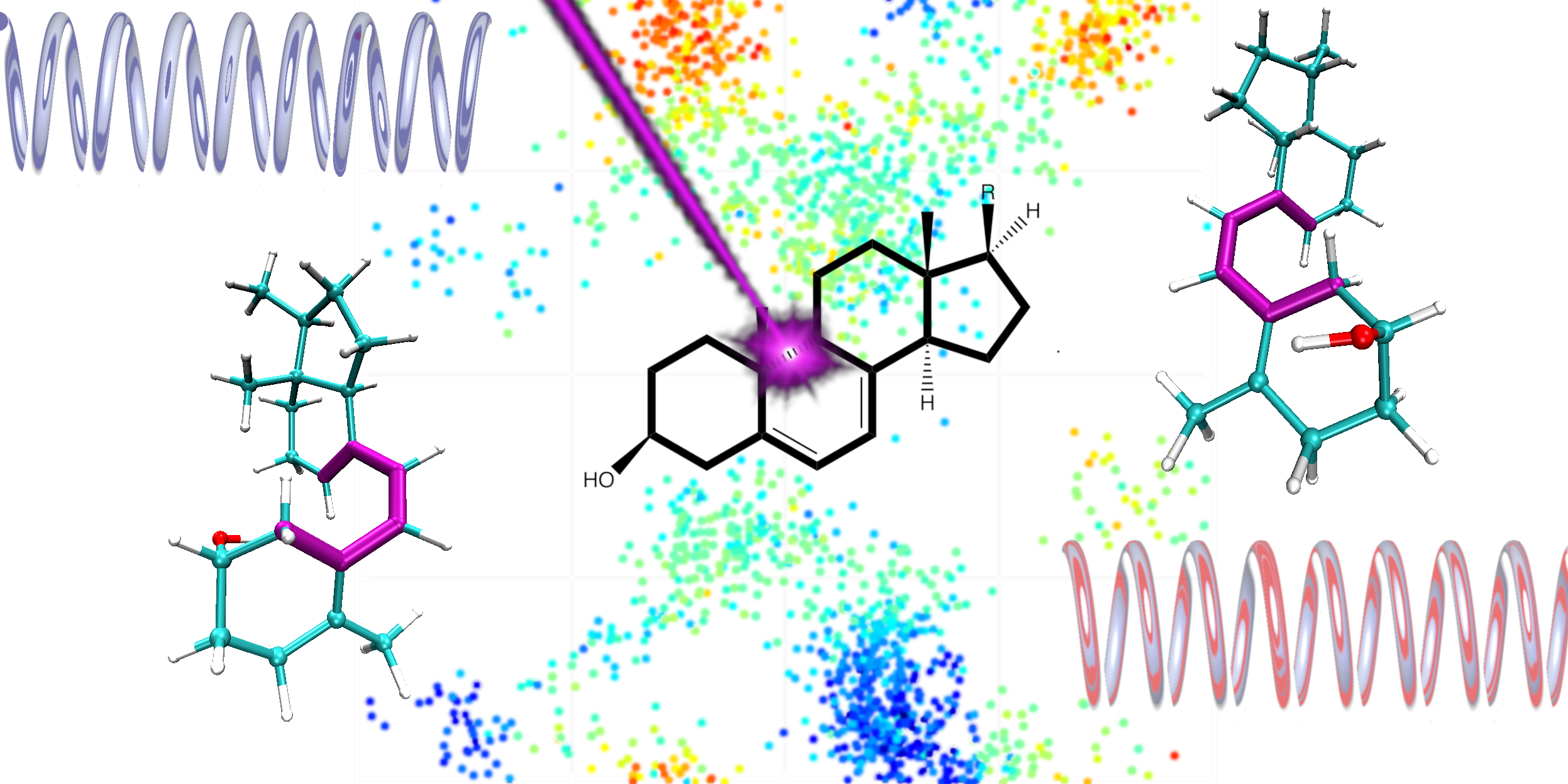}
\end{tocentry}

\maketitle

\begin{abstract}
 We present a method to simulate ultrafast pump-probe time-resolved circular dichroism (TRCD) spectra based on time-dependent density functional theory trajectory surface hopping. The method is applied to simulate the TRCD spectrum along the photoinduced ring-opening of provitamin D.
Simulations reveal that the initial decay of the signal 
is due to excited state relaxation, forming the rotationally flexible previtamin D. 
We further show that oscillations in the experimental TRCD spectrum arise from isomerizations between previtamin D rotamers with different chirality, which are associated with the helical conformation of the triene unit.
We give a detailed description of the formation dynamics of different rotamers, playing a key role in the natural regulation vitamin D photosynthesis.
Going beyond the sole extraction of decay rates, simulations greatly increase the amount of information that can be retrieved from ultrafast TRCD, making it a sensitive tool to unravel details in the sub-picosecond dynamics of photoinduced chirality changes.
\end{abstract}

Ultrafast time-resolved (TR) pump-probe spectroscopy offers the possibility to monitor chemical reactions, such as the making and breaking of bonds, electron transfer, and conformational changes on the femto- to picosecond timescale \cite{zewail2000femtochemistry}. A vast number of different techniques, including TR transient absorption (TA), TR photoelectron ionization, have been developed in the last decades \cite{nuernberger2015multidimensional,
schalk2021photochemistry}.
Conformational changes in chiral molecules can in principal be detected by circular dichroism (CD) and optical rotation spectroscopy.
CD spectroscopy measures the difference in absorption of left- and right-circular polarized light $\Delta \epsilon=\epsilon_L - \epsilon_R$.
Since the electronic CD signal depends on angular relations between the electric and magnetic transition dipole moment,
it is highly sensitive to small changes in the electronic density induced by changes in the three-dimensional structure of chiral molecules.
On the microsecond to second time-scale,
this is a standard tool to study conformational changes in biomolecules, such as proteins and nucleic acids \cite{TRCD_review_1992,goldbeck1997fast}.
However, due to noise caused by density fluctuations of the achiral background, the overall sensitivity of CD is rather small, which makes its application for ultrafast pump-probe spectroscopy challenging \cite{hiramatsu2015communication,MeyerIlse2013}.
Thanks to advances in the experimental set-up, large progress has been made in recent years, leading to pump-probe TRCD spectroscopy with time-resolutions of one picosecond and below \cite{Xiaoliang1989,Niezborala2007,Niezborala2008, trifonov2010broadband,MeyerIlse2012,MeyerIlse2013,hiramatsu2015communication,Steinbacher2015,hache2021multiscale,changenet2022recent}.
While the first ultrafast TRCD measurements with sub-picosecond resolution were restricted to fixed wavelengths \cite{MeyerIlse2012}, recent advances in increasing the sensitivity also allow the broadband measurement of the TRCD \cite{hiramatsu2015communication,oppermann2019ultrafast}.
As common in pump-probe spectroscopy, the TRCD signal can be fitted to decay functions, yielding overall relaxation rates \cite{MeyerIlse2012,mendoncca2013ultrafast}. However, oftentimes the TRCD contains an oscillatory fine structure, which is usually neglected in the analysis, but may potentially give more information about the structural dynamics. 
To obtain a relationship between the oscillatory structure of the TRCD, we apply
non-adiabatic excited state molecular dynamics simulations \cite{Tapavicza2007,Tapavicza2011,Tapavicza2013}, which have been shown to provide structural information of the photodynamics in a variety of organic systems\cite{Wiebeler2014,Oruganti2020} and are therefore well-suited to complement pump-probe experiments \cite{Schalk2016}. 
Here, we apply time-dependent density functional theory surface hopping (TD\-DFT-SH) molecular dynamics simulations to model the TRCD along the photoinduced ring-opening reaction of provitamin D, which constitutes the initial step in natural vitamin D photosynthesis, and for which an experimental TRCD has been measured \cite{MeyerIlse2012}.

\begin{figure}
\centering
  \includegraphics[scale=0.65]{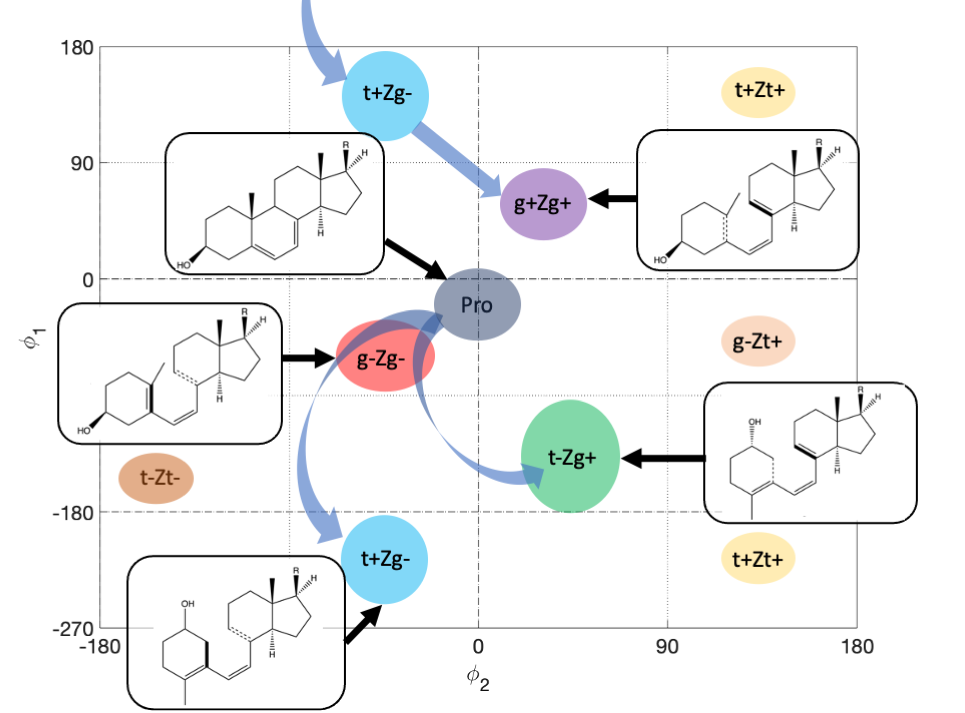}
  \caption{\label{scheme} Photoinduced ring-opening of Pro followed by rotational isomerization of Pre. Dihedral angles $\phi_1$ and $\phi_2$ are defined by the atoms C10-C5-C6-C7 and C6-C7-C8-C9, respectively.}
\end{figure}

The photoinduced electrocyclic ring-opening reaction of 7-Dehydrocholesterol (Figure \ref{scheme}), also known as provitamin D (Pro), has been extensively studied by time-resolved transient absorption spectroscopy in different solvents \cite{Anderson1999,Tang2011} and phospholipid bilayers \cite{sofferman2021ultrafast,Sofferman2021}.
After the ring-opening, which occurs within hundreds of femtoseconds to a few picoseconds after photoexcitation \cite{Tapavicza2011,Snyder2016}, 
the formed rotationally flexible seco-steroid\footnote{seco from Latin secos -- to cut, indicates that the ring of the steroid structures is cut} previtamin D (Pre) undergoes several stages of rotational isomerizations (Figure \ref{scheme}), as non-adiabatic simulations \cite{Tapavicza2011} and TRTA spectroscopy \cite{sofferman2021ultrafast} reveal. 
Rotational isomers of vitamin D seco-steroids are of crucial importance in the intrinsic self-regualtion of vitamin D synthesis\cite{MacLaughlin1982}: On one hand, they give rise to a pronounced wavelength-dependent, conformationally controlled photochemistry  of vitamin D derivatives \cite{Holick1981,MacLaughlin1982,Cisneros2017,Thompson2018,Tapavicza2018}. On the other hand, rotational isomerization possibly affects the thermal formation of vitamin D via a [1,7]-sigmatropic hydrogen shift from C19 to C9, which is thought to be possible only in helical gZg Pre isomers, where hydrogen donor (C19) is in close vicinity to the acceptor atom (C9). This last step in vitamin D formation has been found to be enhanced in biological membranes compared to isotropic solutions, possibly by trapping the gZg conformer due to steric interactions with the phospholipid molecules \cite{holick1995evolutionary,Tian1999}. Understanding the natural vitamin D formation in the skin, therefore requires detailed knowledge about the dynamics and distribution of rotational isomers.
Besides TRTA spectroscopy, the ring-opening reaction in Pro has been investigated by ultrafast time-resolved circular dichroism spectroscopy (TRCD)
by Dietzek and coworkers with a 120~fs time-resolution along fixed probe wavelengths in the UV region.\cite{MeyerIlse2012} Being a biologic paradigm of an ultrafast photoinduced electrocyclic reaction, this reaction also constitutes a perfect test case for ultrafast pump-probe TRCD spectroscopy due to its expected chirality changes on a femtosecond time-scale: The ring-opening causes the molecule to lose one asymmetric carbon center, and furthermore, the central triene unit of Pre can adopt a left- or right-handed helical conformation, two effects that are expected to cause a change in the CD signal. 
The cited study constitutes an important step in the application of ultrafast TRCD and allowed to confirm the previously measured excited state lifetime of Pro of about 1 ps \cite{Fuss1996,Anderson1999,Tang2011}.
However, a closer look at the time-dependent CD signal measured by Dietzek et al.~reveals an oscillatory structure besides the approximate exponential decay. 
In the original work, this feature was not further discussed \cite{MeyerIlse2012}.

To obtain more structural information from TRCD and to determine the cause of the oscillatory structure in the TRCD of Pro, we reinvestigate TDDFT-SH trajectories from our previous study \cite{Tapavicza2013}. 
In TDDFT-SH, 62~\% of the trajectories successfully form the open-ring photoisomer Pre.
The rotational degrees of freedom created by the ring-opening allow the initially formed g-Zg- conformers to relax further, adopting a distribution of different rotamers characterized by the values of $\phi_1$ and $\phi_2$ defined in Fig.~\ref{scheme}. Simulations show that first a rotation around the C5-C6 bond occurs, forming t+Zg- Pre,  which is superimposed by a simultaneous slower rotation around the C7-C8 bond, giving access to other rotamers. Initially isomerization occurs coherently among the trajectories and then increasingly dephase in the motion of the rotational isomerization until they eventually form a Boltzmann ensemble of rotational isomers that does not exhibit any memory of the excited state ring-opening process. The rotational dephasing time amounts to approximately 4--5 ps in gas phase simulations.
In solution, this redistribution process has been found to be almost completed within tens of picoseconds to up to more than 100 ps, depending on the viscosity of the solvent \cite{Anderson1999,Sofferman2021}.

The TDDFT-SH approach is described in detail elsewhere \cite{Tapavicza2013}. 
In brief, initial structures of Pro are obtained from a Boltzmann ensemble at room temperature, generated by Born-Oppenheimer molecular dynamics (BOMD). The nuclear coordinates of the initial structures were propagated using TDDFT nuclear forces of the first singlet excited state (S$_1$). Non-adiabatic coupling vectors between S$_1$ and the ground state (S$_0$) were computed at each timestep and used to compute the {\it fewest switches} probability to non-adiabatic transitions between electronic states according to Tully \cite{Tully1990}.
TDDFT-SH trajectories from our previous study \cite{Tapavicza2011}, which all decayed to S$_0$ within 2 ps of simulation time, were extended by BOMD in S$_0$ to a total simulation time of 4.8 ps to obtain information about the hot ground state dynamics that followed the excited state relaxation.
 
CD spectra can be efficiently calculated by linear response theory \cite{Furche2001,Furche2002,Warnke2012}.
The central quantity of electronic CD is the electric dipole-magnetic dipole polarizability tensor $G_{jk}$. In isotropic systems only its average value
\begin{equation}
G(\omega)=\frac{1}{3}\sum_jG_{jj}(\omega)
\end{equation}
is measured.
The imaginary part of $G(z)$ at real frequency is related to the shape of the CD spectrum. 
Within linear response TDDFT, it can calculated from the response vector $| X, Y\rangle$, resulting from the solution
of the time-dependent Kohn-Sham eigenvalue problem \cite{casida95,Bauernschmitt1996,Furche2001},
\begin{equation}
\text{Im}[G(\omega)]=-\frac{c}{z}\langle \mu^{(j)}|X^{(j)}(z),Y^{(k)}(z)\rangle.
\end{equation}
According to
\begin{eqnarray}
\text{Im}[G(\omega)]& =& 
\frac{c\pi}{3}\sum\frac{1}{\Omega_{0n}}(R_{0n}\delta(\omega-\Omega_{0n})\\\nonumber
& & \,\,\,\,\,\,\,\,\,\,\,\,\,\,\,\,\,\,\,\,\,\,\,\,\,\,\,\,\,\,\,\,\,\,\,\,\,\,\,\,\,\,\,\,\,\,-R_{0n}\delta(\omega+\Omega_{0n}))\, ,  \\\nonumber
\end{eqnarray}
this quantity is related to the rotatory strength
\begin{equation}
R_{0n}=-\text{Im}[{\bf \mu}_{0n} \cdot {\bf m}_{0n}]\, ,
\end{equation}
where ${\bf \mu}_{0n}$ and ${\bf m}_{0n}$ are the electric and magnetic transition dipole moments, respectively.
In practice, the rotatory strength are obtained as standard output from TDDFT calculations \cite{Warnke2012}. Applying Gaussian broadening with a given linewidth (LW) they can be converted to the $\Delta \epsilon$ signal \cite{brown1971electronic}, measured in CD spectroscopy.
To simulate the CD spectra, we averaged CD spectra of single molecular structures obtained from Gaussian broadening of rotatory strengths computed by TDDFT to obtain the macroscopic spectrum of the ensemble of rotational isomers. 

 To compute the TRCD signal, we assume that the CD signal is caused by ground state absorption, rather than excited state absorption.
Provided that the UV pump-pulse induces a $S_1\leftarrow S_0$ transition, 
this is a reasonable assumption if the probe wavelength is also in the UV region,  
since higher $S_n\leftarrow S_1$ absorption energies usually appear at lower energies than the excitation energy of S$_1$. According to this assumption, only the fraction of molecules that have already been relaxed to the ground state after initial excitation gives rise to the CD signal at delay time $\tau$. 
Within surface hopping, the CD signal at time delay time $\tau$ is then calculated as average over the number of trajectories in the ground state ($N_0(\tau)$):
\begin{equation}\label{trcd}
\Delta \epsilon (\tau)=\epsilon_l(\tau)-\epsilon_r(\tau)=1/N\sum_i^{N_0(\tau)}\Delta \epsilon_i(\tau)\, ,
\end{equation}
where $\Delta \epsilon_i(\tau)$ denotes the instantaneous CD spectrum of trajectory $i$ at time $\tau$, obtained from the rotatory strengths of the corresponding molecular structures by Gaussian broadening; $N$ denotes the total number of trajectories.
The $\Delta$CD signal at time $\tau$ is obtained by subtracting the instantaneous spectrum $\Delta \epsilon (\tau)$ from the static CD spectrum of the parent molecule Pro.
In the experimental study \cite{MeyerIlse2012}, however, due to the unknown sign of the instantaneous spectrum, the instantaneous spectrum was added to the static spectrum of Pro and not subtracted. To achieve best comparability between simulated and measured spectrum, we also added the instantaneous spectrum to the static spectrum in our calculation.

\begin{figure}
    \centering
    \includegraphics[scale=0.7]{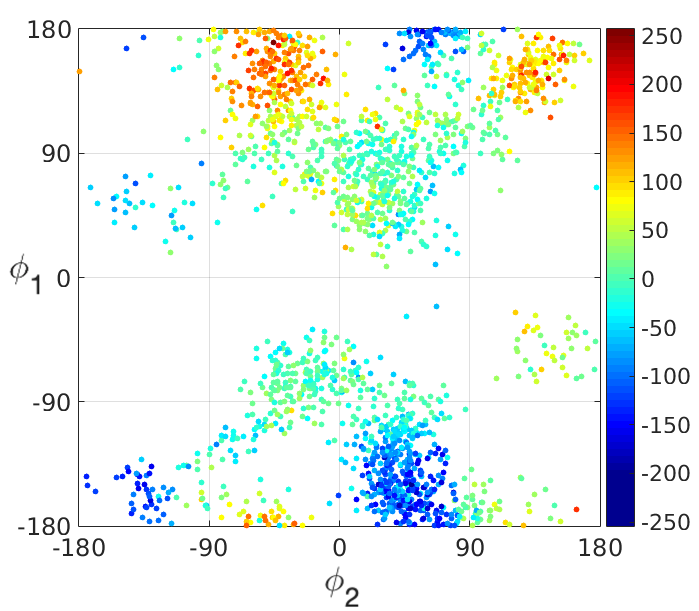}
    \caption{TDDFT rotatory strengths (10$^{-40}$ cgs, length representation) of Pre as function of the dihedral angles $\phi_1$ and $\phi_2$ (defined in Fig.~\ref{scheme}), computed for snapshot structures of Pre generated via REMD. }
    \label{fig:pre_rot}
\end{figure}

\begin{figure}
    \centering
    \includegraphics[scale=0.3]{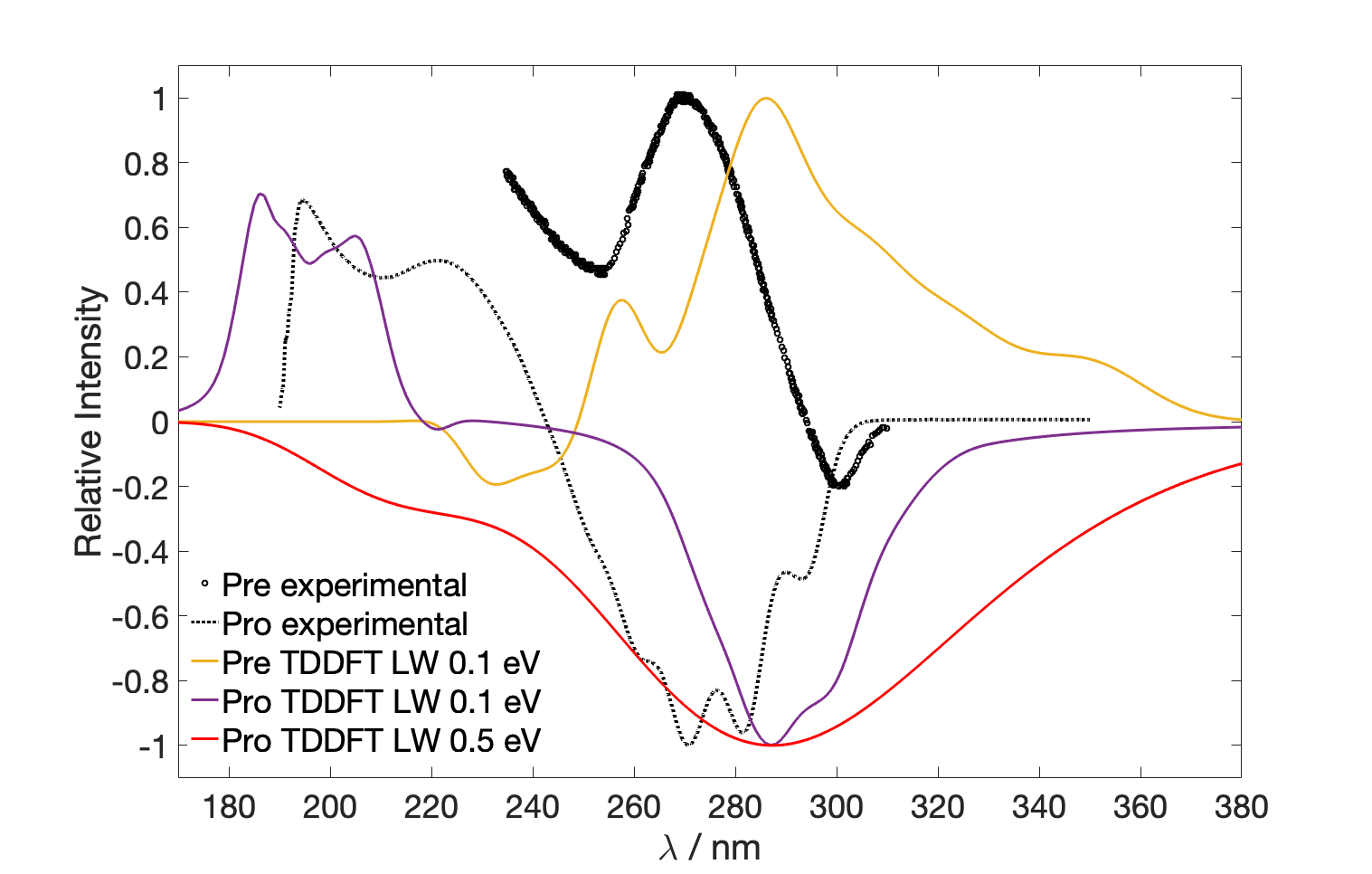}
    \caption{Comparison between static CD spectra of Pro and Pre. Experimental spectrum of Pre from Maessen et al. \cite{maessen1983photochemistry}, measured in ether/iso-pentane/alcohol at 92 K; experimental spectrum of Pro from \cite{MeyerIlse2012}. TDDFT spectra of Pre and Pro at 300 K with a LW of 0.1 eV were computed using the 10 lowest excited states, whereas the TDDFT spectrum of Pro with LW 0.5 eV was computed with lowest 2 excited states. The latter spectrum (red) was used as static reference in the calculation of the TRCD spectrum. All spectra have been normalized to a maximum intensity of 1.}
    \label{static}
\end{figure}

Before we present the TRCD, we assess the dependency of the rotatory strengths on the dihedral angles $\phi_1/\phi_2$ in Pre. To this end, we computed the excitation energies and rotatory strengths for the ground state ensemble of Pre rotamers obtained from replica exchange molecular dynamics (REMD) (Figure \ref{fig:pre_rot}).  From the overall symmetry of this plot, it is visible that the rotatory strengths are sensitive to the dihedral angle conformation of Pre. In particular, the helicality affects the sign and magnitude of the rotatory strengths. Most obviously, this effect emerges in the comparison between the rotatory strengths of t+Zg- and t-Zg+  conformers, which have opposite helicality: values of t+Zg- conformers exhibit positive values ranging from 100--250$\times$10$^{-40}$erg, whereas t-Zg+ conformers exhibit negative rotatory strength of similar magnitude.
A similar opposite relationship appears in the comparison between t+Zt+ and t-Zt- conformers. For gZg conformers, in contrast, rotatory strengths are less sensitive to the dihedral angles, exhibiting values close to zero. 
Since the positive rotatory strengths in the upper half of the plot in Fig.~\ref{fig:pre_rot} are dominating at 300 K, the overall CD spectrum of Pre obtained from Gaussian broadening of the rotatory strengths exhibits a positive band in the 240-380 nm region (Figure~\ref{static}, yellow), which is opposite in sign compared to the static spectrum of Pro (purple). This qualitative trend is confirmed by experimental measurements \cite{maessen1983photochemistry,MeyerIlse2012}. However, both experimental sources do not report the concentration at which the measurements were carried out; therefore, we cannot compare the absolute values of $\Delta \epsilon$ of the molecules. 

\begin{figure}
  \includegraphics[scale=0.23]{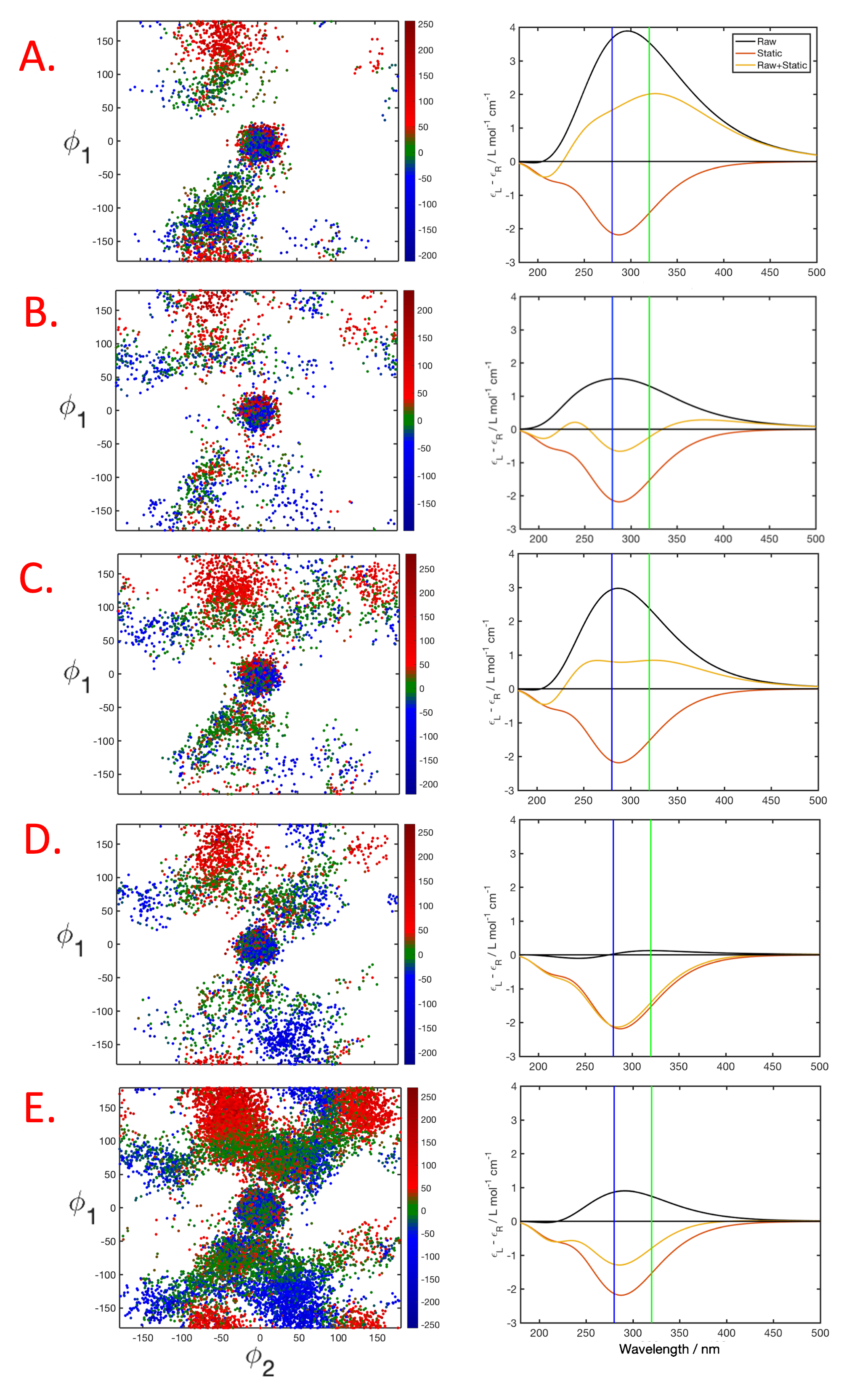}
  \caption{\label{time_windows} 
  Left column: Distribution of structures in the $\phi_1$/$\phi_2$ conformational space for the time windows {\bf A} (633--960 fs), {\bf B} (1180--1392 fs), {\bf C} (1516--1862 fs), {\bf D} (2179--2563 fs), and {\bf E} (3053--4800 fs). Red: positive rotatory strength, blue: negative rotatory strengths, green: rotatory strength close to zero.
  Right column: Instantaneous CD spectrum $\Delta \epsilon$ (black), static spectrum of Pro (red), and difference spectrum $\Delta$CD (yellow), all averaged over the given the time interval. Blue and green vertical lines indicate the wavelengths at which traces were taken to generate Fig.~\ref{broadband}.}
\end{figure}

\begin{figure*}[h]
  \includegraphics[width=\textwidth]{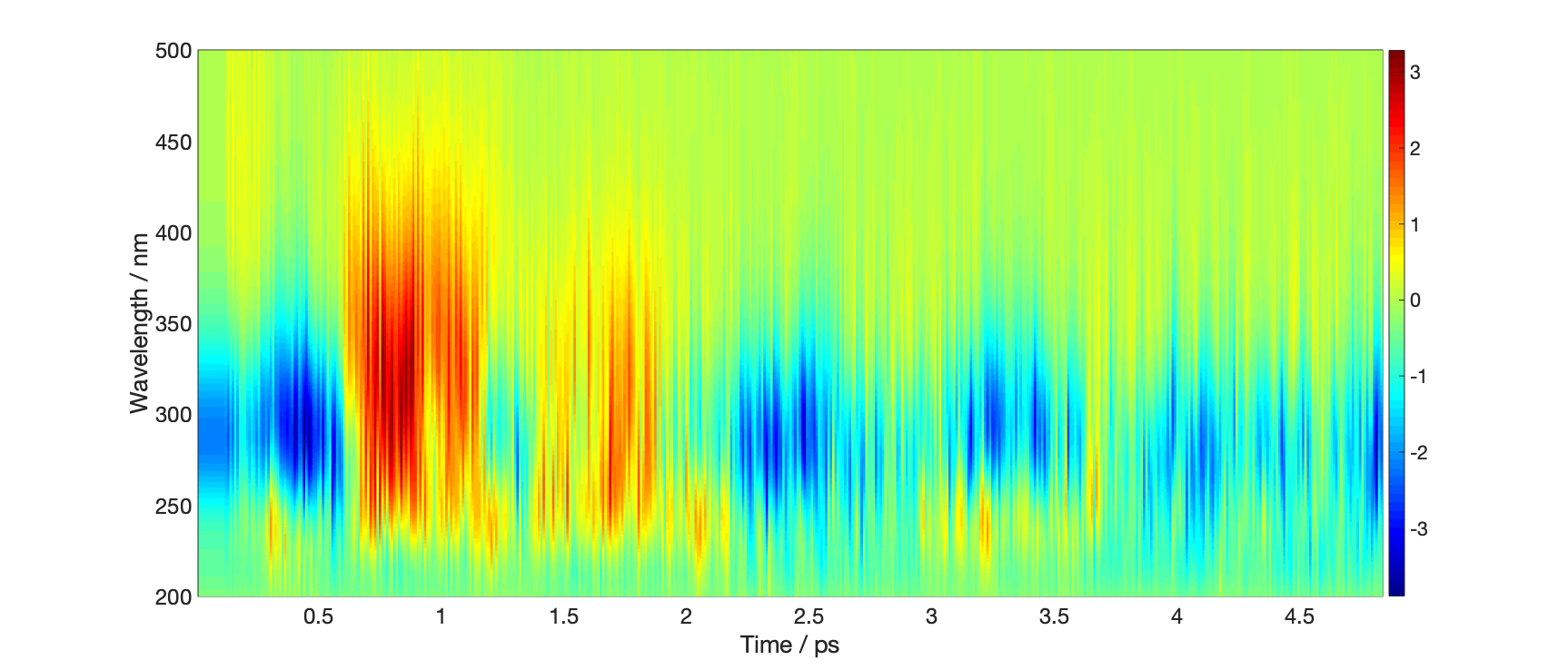}
  \includegraphics[width=\textwidth]{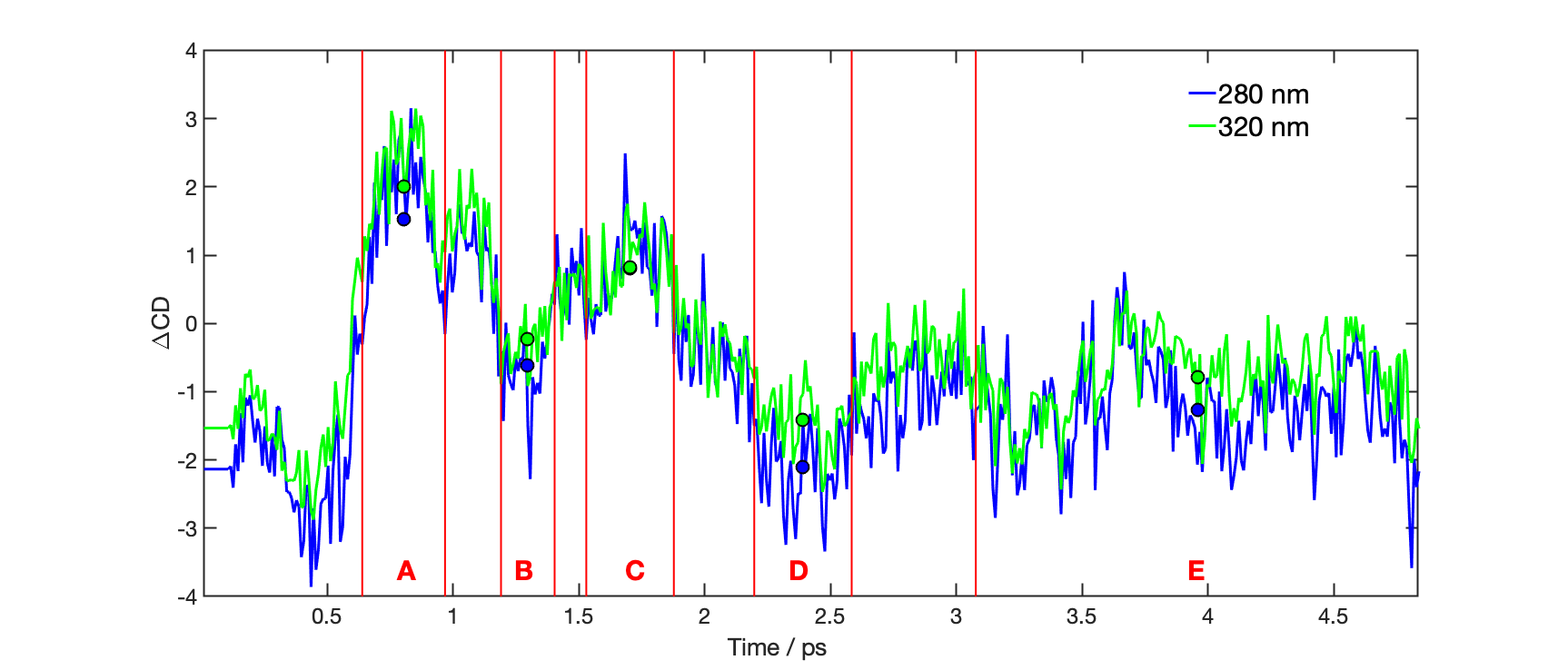}
  \caption{\label{broadband} Upper panel: Simulated broadband time-resolved CD spectrum along the photoinduced Pro ring-opening reaction. The colorbar indicates the $\Delta$CD signal in L/(mol$\times$cm).
 Lower panel: Traces along 280 and 300 nm taken from the broadband spectrum. Blue and green circles indicate the average $\Delta$CD signal over the time windows A, B, C, D, and E, respectively. The distribution of conformers for these time windows is indicated in Fig.~\ref{time_windows}.}
\end{figure*}

To simulate the TRCD, we investigate the dynamics of the first 4.8 picoseconds after initial photoexcitation in terms of the rotatory strengths and the instantaneous CD spectrum (Fig. \ref{time_windows}), allowing us to construct the TRCD spectrum (Fig.~\ref{broadband}). In the time window 600--960~fs (Fig.~\ref{time_windows}, {\bf A}), over 90 \% of the trajectories have decayed to S$_0$ and a substantial amount of the initially formed g-Zg- conformers has isomerized to t+Zg- conformers, exhibiting strongly positive rotatory strengths (Fig.~\ref{time_windows}, {\bf A}, left column), causing an instantaneous CD spectrum with strong intensity and positive sign (Fig.~\ref{time_windows}, {\bf A}, right column, black). 
The strong positive band overcompensates the negative band of the static CD spectrum of Pro (Fig.~\ref{time_windows}, {\bf A}, right culumn, red), leading to a dip in the broadband TRCD spectrum (Fig.~\ref{broadband}, upper panel), as well as in its 280 nm and 320 nm traces (Fig.~\ref{broadband}, lower panel, {\bf A}), that reaches its maximum at approximately 0.8 ps. After reaching this maximum, the TRCD signal partially recovers (Fig.~\ref{broadband}, lower panel, {\bf B}), due to conversion of t+Zg- conformers to g+Zg+ and g+Zt- conformers with low magnitude rotatory strengths (Fig.~\ref{time_windows}, {\bf B}, left), giving rise to a less pronounced positive band in the instantaneous spectrum (Fig.~\ref{time_windows}, {\bf B}, right panel, black). Subsequently, 
the band of the instantaneous spectrum increases due to the formation of t+Zt+ conformers, exhibiting strongly positive rotatory strengths (Fig.~\ref{time_windows}, {\bf C}, left). The following decrease of the instantaneous spectrum is caused by depopulation of t+Zt+ conformers and simultaneous formation of t-Zg+ confomers with negative rotatory strengths (Fig.~\ref{time_windows}, {\bf D}, left). 
The latter oscillation (Fig.~\ref{broadband}, {\bf C} and {\bf D}) appears with lower amplitude than the first one ({\bf A} and {\bf B}), due to an overall dephasing of the ensemble in the $\phi_1/\phi_2$ space until a relatively constant signal is reached (Fig.~\ref{time_windows}, lower panel, {\bf E}). The constant spectrum (Fig.~\ref{time_windows}, {\bf E}, right, yellow) is caused by the superposition of the CD spectra of the equilibrium ensemble of the product Pre, its remaining parent molecule Pro, and the static spectrum of Pro used as reference.
However, since the total simulation time only amounts to 4.8 ps, we cannot ulimately determine if the full equilibrium has been reached.

\begin{figure*}[h]
\end{figure*}

\newpage

Similar oscillations are visible in the experimentally measured TRCD, albeit at lower frequency.
The time in our simulations until the oscillations approximately disappear and a relatively constant TRCD signal is reached ($\approx$ 4 ps) is much shorter than in the experimental spectrum ($\approx$ 14 ps). This difference 
might be caused by two reasons: Most likely the solvent viscosity in the experimental spectrum decelerates the process of rotational isomerization. Furthermore, TDDFT-SH simulations are known to underestimate excited states lifetimes \cite{Granucci2007}, leading to a steeper initial decay of the simulated signal compared to initial decay in the experimental TRCD. 
Nevertheless, our simulations suggest that the oscillations in the experimental spectrum are due to rotational isomerizations, allowing to give a refined picture of this process in solution: 
In solution the first dip is reached at 2--2.5 ps; comparison with
the simulated spectrum allows to assign this time window with formation of the 
high density of t+Zg- structures.
Similarly, the second maximum in the experimental spectrum at 7-8 ps (corresponding to time window {\bf C} in the simulations) indicates  the formation of the t+Zt+ structures.
At approximately 10 ps t-Zg+ conformers are formed, which correspond to time window D in the simulations. 
After about 14 ps, the experimental TRCD appears more or less constant, indicating that an equilibrium ensemble has been reached. However, since the total time in the TRCD measurement amounts to only 18 ps, it cannot be determined if the remaining oscillations are due rotational isomerization or due to noise in the measurements. 

Despite the differences in timescales, the simulations allow to give a detailed description of the isomerization process in solution.
To include solvent viscosity, the simulations could be carried out using QM/MM, adopting a classical desciption of the solvent on a classical level. To achieve better accuracy of the intial decay, more sophisticated non-adiabatic molecular dynamics methods, such as multiple-spawning \cite{ben2000ab} of decoherence corrected surface hopping \cite{granucci2010including} be applied.
Nevertheless, our study shows that TRCD in combination with non-adiabatic molecular dynamics simulations is viable tool to investigate chirality changes on a femto- to picosecond time scale. The synergy between experiment and simulations has the potential to yield sensible information about structural changes that cannot be obtained by the experiment alone. 
\section*{Computational Details}
All DFT and TDDFT calculations employ split valence, triple-$\zeta$ SVP basis sets \cite{Weigend2005} and the hybrid PBE0 \cite{Perdew1996a} functional.
For the non-adiabatic dynamics, the Tully's fewest switches surface hopping \cite{Tully1990} was employed as previously described \cite{Tapavicza2011,Tapavicza2013}. Excited state nuclear gradients non-adiabatic couplings were computed analytically \cite{Furche2002,Send2010}. To prevent imaginary excitation energies TDDFT-SH was applied within the Tamm-Dancoff approximation \cite{Hirata1999}. The nuclear degrees of motion were integrated using the Verlet algorithm \cite{Elliott2000}. 
An NVT ensemble of initial structures and velocities of Pro was generated using BOMD, employing a Nos\'e-Hoover thermostat with a target temperature of 300 K and a characteristic response time of 500 au. 
For BOMD a time step of 50 au was used for the propagation of the nuclear positions, for TDDFT-SH a time step of 40 au was used. More details for the generation of the TDDFT-SH trajectories can be found in our earlier paper \cite{Tapavicza2011}. 
The calculation of the static CD spectrum of Pro and Pre was done using an average of 200 and 500 snapshot geometries, respectively. The lowest ten excitation energies and rotatory strengths. For Pro, BOMD was used to generate the ensemble of structures, whereas for Pre, enhanced sampling using REMD \cite{Sugita1999} was applied, as described elsewhere \cite{Cisneros2017,schalk2021photochemistry,Sofferman2021}. 
For the static spectra, Gaussian line broadening was applied with a LW of 0.1 eV, yielding CD-spectra of the individual snapshot geometries.
For the TRCD spectrum, a LW of 0.5 eV was applied to convert rotatory strengths to $\Delta \epsilon$, for both the static spectrum used as reference, as well as the instantaneous spectrum.
This was necessary to reduce noise due to limited sampling. For computational efficiency, only the lowest two excited states were considered in case of the TRCD. 
The macroscopic spectrum was then calculated as an average of the spectra of the 116 snapshot geometries (Eq. \ref{trcd}) 
at every 10 time steps of 116 TDDFT-SH trajectories. This results in a time resolution of 9.6~fs, which is below the experimental resolution of 120~fs. 
For the TRCD spectra, the full TDDFT response equations were solved. At each of these time steps the CD spectrum was calculated, but only trajectories that have relaxed to the ground state were taken into account. The $\Delta$CD spectrum is usually obtained by subtraction of the instantaneous spectrum from the static spectrum.
However, in the experimental measurement, probe-pulses with opposite circular polarization are alternated using Pockel cells \cite{MeyerIlse2012}, but it is unknown which one of a pair of two consecutive pulses is left-circularly polarized and which one is right-circularly polarized. 
It is thus impossible to gauge the absolute sign of the instantaneous spectrum, therefore it is also possible that the $\Delta$CD spectrum in experimental work\cite{MeyerIlse2012} was computed by adding the instantaneous spectrum to the static spectrum of Pro. We applied both procedures and obtained better agreement with the experimental TRCD spectrum when summation was applied. 
This is also consistent with the fact that the long-time limit of the TRCD is less negative than at time zero: 
Due to the positive sign of the CD spectrum of Pre the long-time limit of the TRCD should be less negative than the static spectrum if addition of the two spectra is applied; this is the case in the experimental spectrum. All electronic structure calculations were performed with Turbomole 6.4 \cite{TURBOMOLE,balasubramani2020turbomole}. The construction of the TRCD was done with MATLAB \cite{MATLAB:2016}.
\begin{acknowledgement}
We acknowledge useful discussions with Benjamin Dietzek and thank him for providing the data from his experiments. Research reported in this paper was supported by National Institute of General Medical Sciences of the National Institutes of Health (NIH) under award numbers R15GM126524, UL1GM118979-02, TL4GM118980, and RL5GM118978. The content is solely the responsibility of the authors and does not necessarily represent the official views of the NIH. We acknowledge technical support from the Division of Information Technology of CSULB.
\end{acknowledgement}


%

\clearpage

\bibliography{papers}

\end{document}